\documentclass{kapedbk}
\usepackage{graphicx}
\normallatexbib
\begin{document}
\articletitle{Entanglement production\\
in a chaotic quantum dot}
\chaptitlerunninghead{Entanglement production in a chaotic quantum dot}
\author{C.W.J. Beenakker, M. Kindermann}
\affil{Instituut-Lorentz, Universiteit Leiden, P.O. Box 9506, 2300 RA
Leiden, The Netherlands}
\author{C.M. Marcus, A. Yacoby\footnote{Visiting from: Department of Condensed
Matter Physics, Weizmann Institute of Science, Rehovot 76100, Israel.}}
\affil{Department of Physics, Harvard University, Cambridge, MA 02138, USA}


\begin{abstract}
It has recently been shown theoretically that elastic scattering in the Fermi
sea produces quantum mechanically entangled states. The mechanism is similar to
entanglement by a beam splitter in optics, but a key distinction is that the
electronic mechanism works even if the source is in local thermal equilibrium.
An experimental realization was proposed using tunneling between two edge
channels in a strong magnetic field. Here we investigate a low-magnetic field
alternative, using multiple scattering in a quantum dot. Two pairs of
single-channel point contacts define a pair of qubits. If the scattering is
chaotic, a universal statistical description of the entanglement production
(quantified by the concurrence) is possible. The mean concurrence turns out to
be almost independent on whether time-reversal symmetry is broken or not. We
show how the concurrence can be extracted from a Bell inequality using
low-frequency noise measurements, without requiring the tunneling assumption of
earlier work.
\smallskip\\
To appear in: {\em Fundamental Problems of Mesoscopic Physics: Interactions and
Decoherence}, edited by I. V. Lerner et al. (Kluwer, Dordrecht, 2004).
\end{abstract}



\section{Introduction}

The usual methods for entanglement production rely on interactions between the
particles and the resulting nonlinearities of their dynamics. A text book
example from optics is parametric down-conversion, which produces a
polarization-entangled Bell pair at frequency $\omega$ out of a single photon
at frequency $2\omega$ \cite{Man95}. In condensed matter the schemes proposed
to entangle electrons make use of the Coulomb interaction or the
superconducting pairing interaction \cite{Egu02}.

Photons can be entangled by means of linear optics, using a beam splitter, but
not if the photon source is in a state of thermal equilibrium
\cite{Sch01,Kim02,Xia02}. Remarkably enough, this optical ``no-go theorem''
does not carry over to electrons: It was discovered recently \cite{Bee03} that
single-particle elastic scattering can create entanglement in an electron
reservoir even if it is in local thermal equilibrium. The existence of a Fermi
sea permits for electrons what is disallowed for photons. The possibility to
entangle electrons without interactions opens up a range of applications in
solid-state quantum information processing \cite{Fao03,Bee03a,Sam03a,Les03}.

Any two-channel conductor containing a localized scatterer can be used to
entangle the outgoing states to the left and right of the scatterer. The
particular implementation described in Ref.\ \cite{Bee03} uses tunneling
between edge channels in the integer quantum Hall effect. In this contribution
we analyze an alternative implementation, using scattering between point
contacts in a quantum dot. We then need to go beyond the tunneling assumption
of Ref.\ \cite{Bee03}, since the transmission eigenvalues $T_{1},T_{2}$ through
the quantum dot need not be $\ll 1$.

The multiple scattering in the quantum dot allows for a statistical treatment
of the entanglement production, using the methods of quantum chaos and
random-matrix theory \cite{Bee97,Mar97,Alh00}. The interplay of quantum chaos
and quantum entanglement has been studied extensively in recent years
\cite{Fur98,Mil99,Zyc01,Ban02,Zni03,Sco03,Jac03}, in the context of
entanglement production by interactions. The interaction-free mechanism studied
here is a new development.

\begin{figure}[tb]
\centerline{\includegraphics[width=10cm]{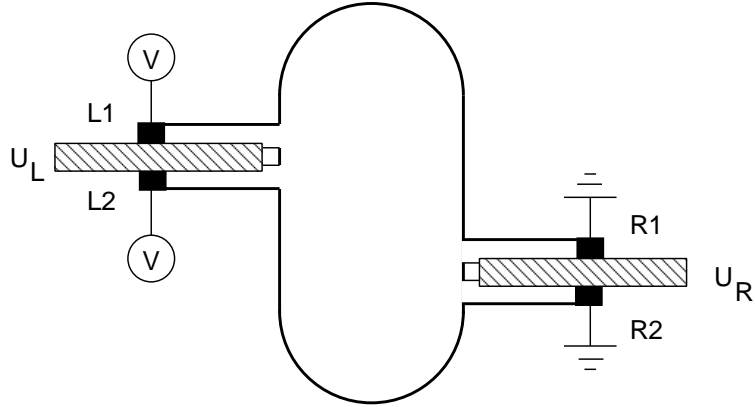}}
\caption{
Sketch of the quantum dot entangler described in the text. An electron leaving
the quantum dot at the left or right represents a qubit, because it can be in
one of two states: it is either in the upper channel (L1,R1) or in the lower
channel (L2,R2). An example of a maximally entangled Bell pair is the
superposition $(|{\rm L1,R1}\rangle+|{\rm L2,R2}\rangle)/\sqrt{2}$.
\label{quantumdotentangler}
}
\end{figure}

The geometry considered is shown in Fig.\ \ref{quantumdotentangler}. A quantum
dot is connected at the left and at the right to an electron reservoir. The
connection is via point contacts connected to single-channel leads. (Spin
degeneracy of the channels is disregarded for simplicity.) There are two leads
at the left (L1, L2) and two leads at the right (R1, R2). A current is passed
through the quantum dot in response to a voltage difference $V$ between the two
reservoirs. We consider the entanglement between the left and right channels in
the energy range $eV$ above the Fermi energy $E_{F}$.

The degree of entanglement is measured through the violation of a Bell
inequality \cite{Cla69} for correlators of current fluctuations
\cite{Cht02,Sam03}. Violation of the Bell inequality requires mixing of the two
outgoing channels at each end of the quantum dot (described by $2\times 2$
unitary matrices $U_{L}$ and $U_{R}$). In order not to modify the degree of
entanglement, this inter-channel scattering should be local, meaning that it
should not lead to backscattering into the quantum dot.\footnote{The mixers
have no effect on the incoming state, because both incoming channels are either
filled or empty at any given energy.}
This might be done by making the barrier that separates lead L1 from L2 (and R1
from R2) partially transparent and tunable by a gate \cite{Sam03} ({\em cf.}
Fig.\ \ref{quantumdotentangler}, shaded rectangles).
\section{Relation between entanglement and transmission eigenvalues}

The incoming state,
\begin{equation}
|\Psi_{\rm
in}\rangle=\prod_{E_{F}<\varepsilon<E_{F}+eV}a_{L,1}^{\dagger}
a_{L,2}^{\dagger}|0\rangle,\label{Psiin}
\end{equation}
factorizes into two occupied channels at the left and two empty channels at the
right, so it is not entangled. Here $a_{L,i}^{\dagger}(\varepsilon)$ is the
creation operator for an incoming excitation at energy $\varepsilon$ in channel
$i$ at the left and $|0\rangle$ represents the Fermi sea at zero temperature
(all states below $E_{F}$ filled, all states above $E_{F}$ empty). There is a
corresponding set of creation operators $a_{R,i}^{\dagger}$ at the right. We
collect the creation operators in a vector
$a^{\dagger}=(a_{L,1}^{\dagger},a_{L,2}^{\dagger},a_{R,1}^{\dagger},
a_{R,2}^{\dagger})$. With this notation we can write the incoming state in the form
\begin{equation}
|\Psi_{\rm in}\rangle=a^{\dagger}\cdot\sigma\cdot a^{\dagger}|0\rangle,\;\;
\sigma=\left(\begin{array}{cc}
(i/2)\sigma_{y}&0\\0&0
\end{array}\right),\label{Psiin2}
\end{equation}
where the product over energies is implicit.

Multiple scattering in the quantum dot entangles the outgoing state in the two
left channels with that in the two right channels. The vector of creation
operators $b^{\dagger}$ for outgoing states is related to that of the incoming
states by a unitary $4\times 4$ scattering matrix: $b=S\cdot a\Leftrightarrow
b^{\dagger}\cdot S=a^{\dagger}$. Therefore the outgoing state has the form
\begin{equation}
|\Psi_{\rm out}\rangle=b^{\dagger}\cdot S\cdot\sigma\cdot S^{T}\cdot
b^{\dagger}|0\rangle.\label{Psiout}
\end{equation}

There are two methods to quantity the degree of entanglement of the outgoing
state:
\begin{itemize}
\item[A.] One can use the entanglement of formation ${\cal F}$ of the full
state $|\Psi_{\rm out}\rangle$. The entanglement of formation of a pure state
is defined by \cite{Nie00}
\begin{equation}
{\cal F}=-{\rm Tr}_{L}\,\rho_{L}\log \rho_{L},\;\;\rho_{L}={\rm
Tr}_{R}\,|\Psi_{\rm out}\rangle\langle\Psi_{\rm out}|,\label{Fdef}
\end{equation}
with ${\rm Tr}_{L}$ or ${\rm Tr}_{R}$ the trace over the degrees of freedom at
the left or right. (The logarithm has base 2.) The entanglement of formation of
the outgoing state is given in terms of the transmission eigenvalues by
\cite{Bee03}
\begin{eqnarray}
{\cal F}&=&-(eV/h)[T_{1}\log T_{1}(1-T_{2})+T_{2}{}\log
T_{2}(1-T_{1})\nonumber\\
&&\mbox{}+(1-T_{1}-T_{2})\log (1-T_{1})(1-T_{2})].\label{Fresult}
\end{eqnarray}
For $T_{1}=T_{2}=1/2$ the rate of entanglement production is maximal, equal to
$2eV/h$ (bits per second).
\item[B.] Alternatively, one can project $|\Psi_{\rm out}\rangle$ onto a state
$|\Psi'_{\rm out}\rangle$ with a single excitation at the left and at the
right, and use the concurrence ${\cal C}$ of this pair of qubits as the measure
of entanglement. The (normalized) projected state is
\begin{equation}
|\Psi'_{\rm out}\rangle=\frac{(1-n_{L,1}n_{L,2})(1-n_{R,1}n_{R,2})|\Psi_{\rm
out}\rangle}{\langle\Psi_{\rm
out}|(1-n_{L,1}n_{L,2})(1-n_{R,1}n_{R,2})|\Psi_{\rm
out}\rangle^{1/2}},\label{Psiprime}
\end{equation}
with number operator $n_{X,i}=b_{X,i}^{\dagger}b_{X,i}^{\vphantom{\dagger}}$
(for $X=L,R$). The concurrence \cite{Woo98} is a dimensionless number between 0
(no entanglement) and 1 (a fully entangled Bell pair).\footnote{
The concurrence ${\cal C}$ of the qubit pair is related to the entanglement of
formation ${\cal F}'$ of the projected state $|\Psi'_{\rm out}\rangle$ by
${\cal F}'=-x\log x-(1-x)\log(1-x)$ with
$x=\frac{1}{2}+\frac{1}{2}\sqrt{1-{\cal C}^{2}}$.}
The transmission formula for the concurrence is \cite{Bee03}
\begin{equation}
{\cal
C}=\frac{2[T_{1}(1-T_{1})T_{2}(1-T_{2})]^{1/2}}
{T_{1}+T_{2}-2T_{1}T_{2}}.\label{Cresult}
\end{equation}
Full entanglement is reached when $T_{1}=T_{2}$, regardless of the value of the
transmission.
\end{itemize}

Notice that in both methods A and B the degree of entanglement depends only on
the transmission eigenvalues $T_{1}$, $T_{2}$, and not on the eigenvectors of
the transmission matrix. Eqs.\ (\ref{Psiprime}) and (\ref{Cresult}) hold
irrespective of whether time-reversal symmetry (TRS) is broken by a magnetic
field or not. In Ref.\ \cite{Bee03} the expressions were simplified by
specializing to the tunneling regime $T_{1},T_{2}\ll 1$. Here we will not make
this tunneling assumption.

In what follows we will concentrate on the concurrence ${\cal C}$ of the
projected state $|\Psi'_{\rm out}\rangle$, since that is the quantity which is
measured by correlating current fluctuations. The entanglement of formation
${\cal F}$ of the full state $|\Psi_{\rm out}\rangle$ contains also
contributions involving a different number of excitations at the left and at
the right. Such contributions are not measurable with detectors that conserve
particle number \cite{Per95}.

\section{Statistics of the concurrence}

The statistics of ${\cal C}$ is determined by the statistics of the
transmission eigenvalues. For chaotic scattering their distribution is given by
random-matrix theory \cite{Bee97},
\begin{equation}
P(T_{1},T_{2})=c_{\beta}|T_{1}-T_{2}|^{\beta}(T_{1}T_{2})^{-1+\beta/2},
\label{PTresult}
\end{equation}
with normalization constants $c_{1}=3/4$, $c_{2}=6$. We obtain the following
values for the mean and variance of the concurrence in the case $\beta=1$
(preserved TRS) and $\beta=2$ (broken TRS):
\begin{eqnarray}
\langle{\cal C}\rangle&=&\left\{\begin{array}{ll}
0.3863&\;\;{\rm if}\;\;\beta=1,\\
0.3875&\;\;{\rm if}\;\;\beta=2,
\end{array}\right.\label{Cmean}\\
\langle{\cal C}^{2}\rangle-\langle{\cal
C}\rangle^{2}&=&\left\{\begin{array}{ll}
0.0782&\;\;{\rm if}\;\;\beta=1,\\
0.0565&\;\;{\rm if}\;\;\beta=2.
\end{array}\right.\label{Cvar}
\end{eqnarray}

The effect of broken TRS on the average concurrence is unusually small, less
than 1\%. In contrast, the conductance $G=(e^{2}/h){\rm
Tr}\,tt^{\dagger}\propto T_{1}+T_{2}$ increases by 25\% upon breaking TRS. The
main effect of breaking TRS is to reduce the sample-to-sample fluctuations in
the concurrence, by about 15\% in the root-mean-square value.

\section{Relation between Bell parameter and concurrence}

The Bell parameter ${\cal E}$ is defined by \cite{Cht02,Sam03}
\begin{equation}
{\cal E}={\rm
max}\,\left[E(U_{L},U_{R})+E(U'_{L},U_{R})+E(U_{L},U'_{R})-E(U'_{L},U'_{R})\right],
\label{Bellparameterdef}
\end{equation}
where the maximization is over the $2\times 2$ unitary matrices
$U_{L},U_{R},U'_{L},U'_{R}$ that mix the channels at the left and right end of
the system. For given $U_{L},U_{R}$ the correlator $E$ has the expression
\begin{equation}
E=\frac{\langle (\delta I_{L,1}-\delta I_{L,2})(\delta I_{R,1}-\delta
I_{R,2})\rangle} {\langle (\delta I_{L,1}+\delta I_{L,2})(\delta I_{R,1}+\delta
I_{R,2})\rangle}.\label{Edef}
\end{equation}
Here $\delta I_{L,i}\equiv I_{L,i}-\langle I_{L,i}\rangle$ is the low-frequency
current fluctuation in the outgoing channel $i$ at the left\footnote{The total
current in channel $i$ at the left (incoming minus outgoing) is
$e^{2}V/h-I_{L,i}$.}
and $\delta I_{R,j}$ is the same quantity for outgoing channel $j$ at the
right. The average $\langle\cdots\rangle$ in this equation is over a long
detection time for a fixed sample. (We will consider ensemble averages later.)

In the tunneling regime $T_{1},T_{2}\ll 1$ there is a one-to-one relation
${\cal E}=2\sqrt{1+{\cal C}^{2}}$ between the Bell parameter ${\cal E}$ and the
concurrence ${\cal C}$. Here we can not make the tunneling assumption. The Bell
parameter (\ref{Bellparameterdef}) can then be larger than expected from the
concurrence. The relation is \cite{Bee03}
\begin{eqnarray}
{\cal E}&=&2\sqrt{1+\kappa^{2}{\cal C}^{2}},\label{EmaxCkappa}\\
\kappa&=&1+\frac{(T_{1}-T_{2})^{2}}{T_{1}(1-T_{1})+T_{2}(1-T_{2})}.
\label{kappadef}
\end{eqnarray}
The amplification factor $\kappa\geq 1$ approaches unity if either
$T_{1}\approx T_{2}$ or $T_{1},T_{2}\ll 1$.

Since ${\cal E}$ gives the amplified concurrence $\kappa{\cal C}$ rather than
the bare concurrence ${\cal C}$, it is of interest to compare the moments of
$\kappa{\cal C}$ with those of ${\cal C}$. By averaging with distribution
(\ref{PTresult}) we find the mean and variance in a chaotic quantum dot:
\begin{eqnarray}
\langle\kappa{\cal C}\rangle&=&\left\{\begin{array}{ll}
0.7247&\;\;{\rm if}\;\;\beta=1,\\
0.8393&\;\;{\rm if}\;\;\beta=2,
\end{array}\right.\label{kappaCmean}\\
\langle\kappa^{2}{\cal C}^{2}\rangle-\langle\kappa{\cal
C}\rangle^{2}&=&\left\{\begin{array}{ll}
0.0838&\;\;{\rm if}\;\;\beta=1,\\
0.0393&\;\;{\rm if}\;\;\beta=2.
\end{array}\right.\label{kappaCvar}
\end{eqnarray}
The amplification by $\kappa$ amounts to about a factor of two on average.

\section{Relation between noise correlator and concurrence}

For a different perspective on the relation between noise and entanglement, we
write the correlator (\ref{Edef}) of current fluctuations in a form that
exposes the contribution from the concurrence.

Low-frequency correlators can be calculated with the help of the formula
\cite{But90}
\begin{equation}
\lim_{\omega,\omega'\rightarrow 0}\langle\delta I_{L,i}(\omega)\delta
I_{R,j}(\omega')\rangle=-(e^{3}V/h)2\pi\delta(\omega+\omega')
|(rt^{\dagger})_{ij}|^{2}.\label{correlator}
\end{equation}
The reflection and transmission matrices $r,t$ are to be evaluated at the Fermi
energy. We decompose these matrices in eigenvectors and eigenvalues,
\begin{equation}
r=U_{L}\left(\begin{array}{cc}
\sqrt{1-T_{1}}&0\\0&\sqrt{1-T_{2}}
\end{array}\right)U_{0},\;\;
t=U_{R}\left(\begin{array}{cc}
\sqrt{T_{1}}&0\\0&\sqrt{T_{2}}
\end{array}\right)U_{0},\label{decomposition}
\end{equation}
with $2\times 2$ unitary matrices $U_{L},U_{R},U_{0}$. The matrix $r$ contains
the reflection amplitudes from left to left and the matrix $t$ contains the
transmission amplitudes from left to right.\footnote{In the presence of TRS one
has $U_{0}=U_{L}^{T}$, but this constraint is irrelevant because anyway $U_{0}$
drops out of Eq.\ (\ref{correlator}).}

Substitution into Eq.\ (\ref{Edef}) gives
\begin{equation}
E=(1-2|U_{L,11}|^{2})(1-2|U_{R,11}|^{2})+4\kappa{\cal C}\,{\rm
Re}\,U^{\vphantom{\ast}}_{L,11}U^{\ast}_{R,11}U^{\ast}_{L,12}
U^{\vphantom{\ast}}_{R,12}.
\label{ECrelation}
\end{equation}
We see that the entire dependence of the correlator $E$ on the transmission
eigenvalues is through the product $\kappa{\cal C}$ of concurrence and
amplification factor. This is the same quantity that enters in the Bell
parameter (\ref{EmaxCkappa}). The correlator $E$ is less useful for the
detection of entanglement than the Bell parameter ${\cal E}$, because it
depends also on the matrices of eigenvectors $U_{L},U_{R}$ --- which the Bell
parameter does not.

In a chaotic quantum dot the two matrices $U_{L}$ and $U_{R}$ are independently
distributed in the circular unitary ensemble (a socalled ``isotropic''
distribution \cite{Bee97}). Averages over these matrices can be done
conveniently in the parametrization
\begin{eqnarray}
&&U=e^{i\alpha}\left(\begin{array}{cc}
e^{i\phi}\cos\gamma&e^{i\psi}\sin\gamma\\
-e^{-i\psi}\sin\gamma&e^{-i\phi}\cos\gamma
\end{array}\right),\label{UXdef}\\
&&\gamma\in(0,\pi/2),\;\;\alpha,\phi,\psi\in(0,2\pi).\label{angles}
\end{eqnarray}
The isotropic distribution implies that all four angles
$\gamma,\alpha,\phi,\psi$ are independent. The distribution of
$\alpha,\phi,\psi$ is uniform while the distribution of $\gamma$ is
$P(\gamma)\propto\sin 2\gamma$.

\begin{figure}[tb]
\centerline{\includegraphics[width=10cm]{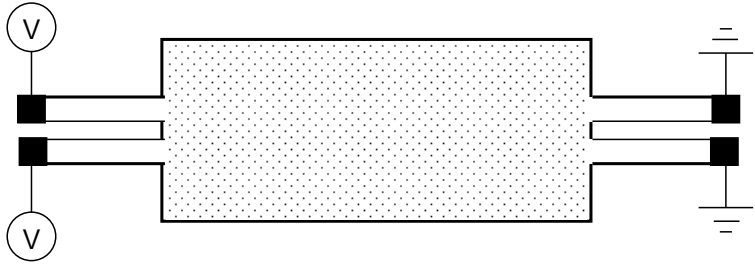}}
\caption{
The quantum dot of Fig.\ \protect\ref{quantumdotentangler} has been replaced by
a disordered wire (dotted rectangle). Although the distribution of transmission
eigenvalues is different, the relation (\protect\ref{Evariance}) between noise
correlator and concurrence still applies. This relation only relies on the
isotropy of the eigenvector distribution.
\label{wireentangler}
}
\end{figure}

With this parametrization Eq.\ (\ref{ECrelation}) takes the form
\begin{equation}
E=\cos 2\gamma_{L}\cos 2\gamma_{R}+\kappa{\cal C}\sin 2\gamma_{L}\sin
2\gamma_{R}\cos(\phi_{L}-\psi_{L}-\phi_{R}+\psi_{R}).\label{Eangles}
\end{equation}
Upon averaging over the angles we find
\begin{equation}
\langle E\rangle=0,\;\;\langle
E^{2}\rangle=\frac{1}{9}+\frac{2}{9}\langle\kappa^{2}{\cal C}^{2}\rangle.
\label{Evariance}
\end{equation}
The significance of this equation is that it applies generally to $2\times 2$
transmission matrices with an isotropic distribution of eigenvectors, even if
the distribution of eigenvalues differs from Eq.\ (\ref{PTresult}). For
example, it applies to the disordered conductor shown in Fig.\
\ref{wireentangler}.

\section{Bell inequality without tunneling assumption}

The Bell parameter ${\cal E}$ is no longer directly related to the concurrence
${\cal C}$ if the transmission probabilities are not small compared to unity
\cite{Bee03}: The relation (\ref{EmaxCkappa}) between ${\cal E}$ and ${\cal C}$
contains a spurious amplification factor $\kappa\geq 1$, which approaches unity
in the tunneling regime. The same amplification factor appears in the
correlator (\ref{ECrelation}). In this section we show how one can avoid the
amplification factor, by calculating the violation of the Bell inequality
without making the tunneling assumption. The same problem was studied, from a
different perspective, in Ref.\ \cite{Sam03a}.

As described in Ref.\ \cite{Cht02}, the Bell inequality is formulated in terms
of the correlator $K_{ij}$ of the number of outgoing electrons detected in a
time $\tau$ in channel $i$ at the left and channel $j$ at the right:
\begin{eqnarray} K_{ij}&=&\tau^{-2}\int_{0}^{\tau}dt\int_{0}^{\tau}dt'\,\langle
I_{L,i}(t)I_{R,j}(t')\rangle\nonumber\\
&=&\langle I_{L,i}\rangle\langle
I_{R,j}\rangle+\int_{-\infty}^{\infty}d\omega\,
\frac{2\sin^{2}(\omega\tau/2)}{\pi(\omega\tau)^{2}}C_{ij}(\omega).
\label{Bellcorrelator}
\end{eqnarray}
Here $C_{ij}(\omega)$ is the frequency dependent correlator of current
fluctuations,
\begin{equation}
C_{ij}(\omega)=\int_{-\infty}^{\infty}dt\,e^{i\omega t}\langle\delta
I_{L,i}(t)\delta I_{R,j}(0)\rangle.\label{Cijdef}
\end{equation}
In the tunneling limit it is possible to neglect the product of averages
$\langle I_{L,i}\rangle\langle I_{R,j}\rangle$ and retain only the second term
in Eq.\ (\ref{Bellcorrelator}), proportional to the current correlator
$C_{ij}$. Both terms are needed if one is not in the tunneling limit.

We assume that $V$ is small enough that the energy dependence of the scattering
matrix may be neglected in the range $(E_{F},E_{F}+eV)$. (That requires $eV$
small compared to the mean level spacing of the quantum dot.) Then the
frequency dependence of $C_{ij}(\omega)$ is given simply by\footnote{The
cross-correlator $C_{ij}(\omega)$ vanishes for $|\hbar\omega|>eV$ because we
are correlating only the outgoing currents; the correlator of incoming plus
outgoing currents contains also a voltage-independent term
$\propto|\hbar\omega|$, cf.\ Ref.\ \protect\cite{Bla00}.}
\begin{equation}
C_{ij}(\omega)=C_{ij}(0)\times\left\{\begin{array}{cl}
1-|\hbar\omega/eV|&\;\;{\rm if}\;\;|\hbar\omega/eV|<1,\\
0&\;\;{\rm if}\;\;|\hbar\omega/eV|>1.
\end{array}\right.\label{Cijomegaresult}
\end{equation}
For short detection times $\tau\ll h/eV$ one may take the limit
\begin{equation}
\lim_{\tau\rightarrow
0}\int_{-\infty}^{\infty}d\omega\,\frac{2\sin^{2}(\omega\tau/2)}
{\pi(\omega\tau)^{2}}C_{ij}(\omega)=\int_{-\infty}^{\infty}
\frac{d\omega}{2\pi}C_{ij}(\omega)= \frac{eV}{h}C_{ij}(0).\label{Cijzero}
\end{equation}

In view of Eq.\ (\ref{correlator}), the zero-frequency limit of the current
correlator (\ref{Cijdef}) is given by
\begin{equation}
C_{ij}(0)=-(e^{3}V/h)|(rt^{\dagger})_{ij}|^{2}.\label{Cijzeroresult}
\end{equation}
The mean outgoing currents are given by $\langle
I_{L,i}\rangle=(e^{2}V/h)(rr^{\dagger})_{ii}$ and $\langle
I_{R,j}\rangle=(e^{2}V/h)(tt^{\dagger})_{jj}$. Substitution into Eq.\
(\ref{Bellcorrelator}) gives the short-detection-time limit
\begin{eqnarray}
\lim_{\tau\rightarrow 0}K_{ij}&=&\langle I_{L,i}\rangle\langle
I_{R,j}\rangle+(eV/h)C_{ij}(0)\nonumber\\
&=&(e^2
V/h)^{2}\left[(rr^{\dagger})_{ii}(tt^{\dagger})_{jj}-
|(rt^{\dagger})_{ij}|^{2}\right].\label{Bellcorrelatorzero}
\end{eqnarray}

We now define the correlator $\tilde{E}$ in terms of the short-time $K_{ij}$,
\begin{equation}
\tilde{E}=\frac{K_{11}+K_{22}-K_{12}-K_{21}}{K_{11}+K_{22}+K_{12}+K_{21}}.
\label{Etildedef}
\end{equation}
Notice that this definition of $\tilde{E}$ corresponds to definition
(\ref{Edef}) of $E$ if $K_{ij}$ is replaced by $C_{ij}(0)$. Substitution of
Eq.\ (\ref{Bellcorrelatorzero}) leads to
\begin{equation}
\tilde{E}=-(1-2|U_{L,11}|^{2})(1-2|U_{R,11}|^{2})-4{\cal C}\,{\rm
Re}\,U^{\vphantom{\ast}}_{L,11}U^{\ast}_{R,11}U^{\ast}_{L,12}
U^{\vphantom{\ast}}_{R,12},\label{EtildeCrelation}
\end{equation}
where we have used the parametrization (\ref{decomposition}). Apart from an
overall minus sign, Eq.\ (\ref{EtildeCrelation}) is the same as Eq.\
(\ref{ECrelation})
--- but without the factor $\kappa$ multiplying the concurrence.

The maximal violation $\tilde{\cal E}$ of the Bell inequality is defined in the
same way as in Eq.\ (\ref{Bellparameterdef}), with $E$ replaced by $\tilde{E}$.
The result
\begin{equation}
\tilde{\cal E}=2\sqrt{1+{\cal C}^{2}}\label{tildeEmax}
\end{equation}
is the same as Eq.\ (\ref{EmaxCkappa}) --- but now without the factor $\kappa$.

Since short-time detection experiments are very difficult in the solid state,
the usefulness of Eq.\ (\ref{tildeEmax}) is that it allows one to determine the
concurrence using only low-frequency measurements. It generalizes the result of
Ref.\ \cite{Sam03} to systems that are not in the tunneling regime and solves a
problem posed in Ref.\ \cite{Bee03} (footnote 24).

\section{Conclusion}

We have investigated theoretically the production and detection of entanglement
by single-electron chaotic scattering. Much is similar to the tunneling regime
studied earlier \cite{Bee03}, but there are some interesting new aspects:
\begin{itemize}
\item The degree of entanglement, quantified by the concurrence ${\cal C}$, is
sample specific. The sample-to-sample fluctuations become smaller if
time-reversal symmetry is broken, while the average concurrence is almost
unchanged.
\item The low-frequency current correlator $C_{ij}$ and the Bell parameter
${\cal E}$ constructed from it give the concurrence times an amplification
factor $\kappa$. In the tunneling regime $\kappa\rightarrow 1$. One also has
$\kappa=1$ if the two transmission eigenvalues $T_{1},T_{2}$ are equal. The
factor $\kappa$ can become arbitrarily large if $T_{1}\rightarrow 1$ and
$T_{2}\rightarrow 0$ (or vice versa). On average, the amplification factor in
an ensemble of chaotic quantum dots is about a factor of two.
\item The bare concurrence, without the amplification factor, is obtained by
adding to the low-frequency current correlator the product of average currents
times $h/eV$.
\item The concurrence gives the maximal violation of the Bell inequality for
detection times $\tau$ short compared to the coherence time $h/eV$. In Ref.\
\cite{Cht02} the opposite limit $\tau\gg h/eV$ was taken, which is appropriate
in the tunneling regime, but  does not allow to violate the Bell inequality
outside of that regime. A similar conclusion was reached in Ref.\
\cite{Sam03a}.
\end{itemize}

{}From an experimental point of view, the missing building block in Fig.\
\ref{quantumdotentangler} is the local mixer at the left and right end of the
quantum dot. These mixers are needed to isolate the contribution to the noise
correlator that is due to the concurrence. In optics, a simple rotation of the
polarizer suffices. The electronic analogue is a major challenge.

\begin{acknowledgments}
We have benefitted from comments on a draft of this manuscript by Markus
B\"{u}ttiker, Peter Samuelsson, and Eugene Sukhorukov. This research was
supported by the Dutch Science Foundation NWO/FOM,
by the U.S. Army Research Office (Grants DAAD 19-02-1-0086 \&
DAAD-19-99-1-0215), and by the Harvard University CIMS Visitors Program.
\end{acknowledgments}

\end{document}